# Impact of COVID-19 Lockdown Measures on Chinese Startups and Local Government Public Finance: Challenges and Policy Implications

Xin Sun
Georgia Institute of Technology

Advisor:
Daniel Dench
Georgia Institute of Technology

April 2023

# Contents

*1.Introduction*

This paper aims to assess the impact of COVID-19 on the public finance of Chinese local governments, with a particular focus on the effect of lockdown measures on startups during the pandemic. The outbreak has placed significant fiscal pressure on local governments, as containment measures have led to declines in revenue and increased expenses related to public health and social welfare. In tandem, startups have faced substantial challenges, including reduced funding and profitability, due to the negative impact of lockdown measures on entrepreneurship. Moreover, the pandemic has generated short- and long-term economic shocks, affecting both employment and economic recovery. To address these challenges, policymakers must balance health concerns with economic development. In this regard, the government should consider implementing more preferential policies that focus on startups to ensure their survival and growth. Such policies may include financial assistance, tax incentives, and regulatory flexibility to foster innovation and entrepreneurship. By and large, the COVID-19 pandemic has had a profound impact on both the public finance of Chinese local governments and the startup ecosystem. Addressing the challenges faced by local governments and startups will require a comprehensive approach that balances health and economic considerations and includes targeted policies to support entrepreneurship and innovation.

*1.1 Introduction of the impact of lockdown on startups in China during COVID-19*

The COVID-19 pandemic, which emerged in late 2019, has swept the globe with unprecedented speed, impacting the health and livelihoods of millions of people worldwide. While the primary focus has been on the pandemic's public health effects, it has also become increasingly apparent that COVID-19 poses significant economic challenges for countries around the world. Governments worldwide have had to impose stringent measures to contain the spread of the virus, which has led to widespread disruptions in economic activity. These measures have included restrictions on people's movements, the closure of non-essential businesses, and restrictions on international trade. Such measures have severely impacted global supply chains, limiting the allocation of resources and exacerbating pre-existing global economic challenges. The lockdown measures implemented by many countries have had a particularly severe impact on the production and employment of numerous industries, particularly those that require physical interaction, such as tourism and transportation. These industries have suffered significant revenue losses due to travel restrictions and lower demand. The disruptions to these industries have also had a knock-on effect on the broader economy, leading to increased levels of unemployment and reducing overall economic growth.

In China, the government successfully brought the COVID-19 pandemic under control by April 2020 and shifted its focus towards preventing imported cases and asymptomatic infections. However, despite the control of the pandemic, the crisis resulted in several unavoidable outcomes. Firstly, export-oriented firms faced significant challenges due to the widespread nature of the pandemic, causing obstacles in supply chains and disruptions in production. This led to significant financial difficulties for many firms. Secondly, due to the unpredictable nature of the pandemic, it was challenging to determine when it would be safe to resume work and production. Although the government allowed some factories to resume operations, any instance of illness resulted in temporary shutdowns, leading to discontinuity in production and bankruptcy for many firms. Thirdly, enterprises faced significant operational risks due to supply chain and capital chain disruptions, which were formidable challenges for businesses during the pandemic. The sharp drop in orders and cost pressures further intensified these risks. Fourthly, the pandemic led to changes in consumer behavior, with a significant shift in demand towards contactless services and essential goods. This created an emergent need for a responsive supply system, which presented challenges for startups in terms of funding and profitability. Overall, these outcomes illustrate the multidimensional challenges presented by the COVID-19 pandemic and have significant implications for businesses and policymakers alike.

Start-ups face significant challenges and vulnerabilities in competing in the market. They frequently encounter issues such as insufficient funding, talent shortages, and difficulties in developing their businesses. The COVID-19 pandemic has further amplified these issues, endangering the survival of many start-ups. The pandemic has caused widespread production disruptions, depressed domestic consumption, and inflexible expenses related to rents, wages, and interest expenses, all of which have put immense strain on start-ups' fragile capital chains. As a result, a substantial number of start-ups have faced bankruptcy. The pandemic has had profound implications for the start-up ecosystem, highlighting the importance of robust and adaptable business models and adequate resources for start-up survival.

This paper aims to examine the impact of the COVID-19 pandemic on the number of new startups in China. Specifically, we use monthly province-level data to measure the number of startups based on the number of new registered enterprises in all 31 provinces in China. Through this research, we seek to determine whether the lockdown measures imposed during the pandemic had a significant effect on the establishment of new startups in China. Additionally, we aim to provide empirical evidence on the development of entrepreneurship in China.

The economic ramifications of COVID-19 have been extensively studied by researchers, who widely agree that the pandemic has significantly impacted the global economy. COVID-19 has led to reduced demand, factory closures, negative sales gaps, and long-lasting scars. Policymakers face a difficult trade-off between preserving public health and maintaining income and employment, while also potentially limiting mobility. Lockdown policies have resulted in reduced income and expenditure for residents, with their spending serving as a direct source of income for other residents and enterprises. In China, the rapid spread of COVID-19 has disrupted normal life, with the potential for long-term effects on the country's economy. The intensification of the pandemic may have significant implications for China's economic future.

COVID-19 has resulted in both short-term and long-term economic shocks. In the short term, China's lockdown policies, which include isolation and social distancing measures, have reduced domestic market demand and led to unemployment. Research focusing on Chinese agriculture found that the sector suffered due to a decrease in global demand for Chinese agricultural products in 2020. Looking ahead to the long term, the demand shock induced by lockdown policies in France is believed to be temporary, with the economy potentially recovering towards its baseline trajectory over the following decade.

In recent times, the level of entrepreneurship in developing countries has been linked to the wealth and poverty of these nations, yet it remains one of the least studied significant economic and social phenomena. Researchers studying entrepreneurship and startups have focused on corporate social responsibility of Small and Medium-Sized Enterprises (SMEs), which offers an explanation for the lack of responsible entrepreneurship among SMEs and lists obstacles to entrepreneurship in developing countries. With regards to the pandemic period, estimates suggest that the disruptions caused by COVID-19 may have long-lasting effects on employment, with the potential for substantial losses lasting over a decade even if the slump in startup activity is only temporary. Furthermore, lockdown policies have had a negative impact on firm creation, which has significant policy relevance.

*1.2 Introduction of the impact of COVID-19 on local government public finance in China*

The rapid spread of the Covid-19 epidemic has caused major economies to suffer and led to stagnating global economic growth. China's economic and social development has also been seriously impacted. From 2011 to 2020, the scale of local fiscal expenditures in China has been larger than the scale of revenues, leading to a widening fiscal balance year by year. Since 2015, the growth rate of fiscal expenditure has exceeded the growth rate of revenue, aggravating the contradiction between revenue and expenditure. Although local public budget expenditure in 2020 increased by 3.36% year-on-year to 210.583 billion yuan.

According to fiscal data released by the Ministry of Finance by the end of 2022, China's economy experienced mixed results with national public budget revenue reaching 20,370,348 million yuan, up 0.6% from 2021. National public budget expenditures totaled 26,194,480 million yuan, an increase of 6.1%. Meanwhile, China's local government debt balance rose to 16.47 trillion yuan from 15.32 trillion yuan in 2021, up 7.51% year-on-year. Of this amount, general debt accounts for 62.70% and special local government debt accounts for 37.30%.

The impact of the epidemic in 2020 has led to a faster growth rate of health affairs expenses in each local government's fiscal expenditures. Due to the decline in local tax revenue caused by the sequestration, fiscal revenue is less robust, and the sustainability of the debt decreases. In addition, some local governments are dependent on land finance, with concessions of state-owned land use rights becoming the main source of government fund revenue.

Public finance is the foundation and important guarantee of nation-building and governance. It not only reflects the national economic situation comprehensively but also lays the foundation for the national government to carry out macro-control of the market economy. Fiscal revenue is an important indicator to measure the financial status of local governments. The abundance and stability of fiscal revenue determine the ability of local governments to provide public goods and services in economic and social activities. The financial relationship between central and local government is essential for understanding the composition of local government revenue and expenditure.

Central-local fiscal relations generally include three aspects: first, the division of revenue, which mainly refers to how government revenues, mainly from taxes, are allocated between the central and local governments. Taxes are usually categorized according to their attributes, with taxes that reflect national sovereignty being classified as central government revenue, such as customs duties and import taxes, while taxes with a relatively stable tax base and clear regional attributes are classified as local government revenue, such as property taxes and municipal taxes. Taxes with strong liquidity and uneven distribution between regions are classified as central government revenue or shared revenue, with the central government usually having a higher share.

Second, the division of fiscal powers and expenditure responsibilities is based on three principles. The first is the principle of scope of benefits, which means that if a certain expenditure affects areas outside the region and has certain externalities, it should be borne by a higher level of government. The second is the principle of complexity of information, which means that the more complex the information acquisition and processing is, and the more likely it is to cause information asymmetry, the more it should be the responsibility of local and especially grassroots governments. The third is the principle of stimulating initiative, which means that the division of fiscal powers should fully reflect the matching of rights and responsibilities, which is conducive to the active performance of duties and incentives and achieving overall maximization of benefits.

Third, the transfer payment system typically includes two types of payments: general transfer payments and special transfer payments. The former is an unconditional grant, while the latter must be used for designated purposes. General transfer payments are primarily used to subsidize local affairs, whereas special transfer payments are mainly employed for centrally entrusted affairs, central and local joint affairs, and local affairs requiring guidance and encouragement. Additionally, there are some transfer payment models that lie between general and special transfer payments, such as the classification transfer payment model for central and local joint affairs involving basic livelihoods, like education and health. In general, local government revenue mainly comes from two sources: on one hand, the income of party-owned enterprises and various tax revenues, which include business tax, local enterprise income tax, personal income tax, urban land use tax, and land value-added tax; on the other hand, there are fiscal transfer subsidies from the central government.

Fiscal expenditures are primarily divided into education, medical, and social security expenditures, reflecting the government's investment in constructing basic livelihood protection. Total financial expenditure mainly consists of capital expenditure, science and technology support costs, industrial transportation business expenses, and health business expenses, among others.

Local government debt refers to the debt raised or guaranteed by the local government as the debtor in the name of the government, and the resulting obligation to repay the funds. It is issued as a form of revenue-raising for local governments, and its revenue is included in local government budgets and arranged and dispatched by local governments. It is an essential means for local governments to raise funds to cover their fiscal deficits. By 2013, the total outstanding local government debt as a percentage of provincial revenue exceeded 100 percent for all provinces, and in some poorer provinces, the ratio was as high as 500 percent (Adam Y. Liu et al. 2022).

Since the Tax Sharing Reform in 1994, the local government revenue of the People's Republic of China (PRC) has faced downward risk problems (Ziying Fan et al. 2007). With the outbreak of Covid-19 in 2020, local governments in China have implemented non-pharmaceutical interventions, such as isolating confirmed patients and their close contacts, closing public places like schools, limiting or stopping market gatherings, restricting transportation, and issuing domestic and international travel bans. Although containment measures and the extent of the epidemic varied across cities, local fiscal pressure increased from 0.38 to 0.435, a rise of 14.53%, and the fiscal revenue and expenditure gap accounted for almost half of the local fiscal revenue (Yue Mei Guo et al. 2021). The lockdown policy directly or indirectly affected local government revenue capacity, and this paper will analyze the impact of different urban outbreak indicators on local government finances.

*2. Hypothesis Development*

*2.1 Hypothesis Development* of *the Impact of Lockdown on Startups*

Drawing upon a literature review and identifying a research gap, this study aims to develop hypotheses H1a, H1b, and H1c to test the effect of lockdown measures on startups in China during the COVID-19 pandemic. The primary objective is to investigate the relationship between the stringency of lockdown measures and the number of newly registered startups. Specifically, this study aims to test the following hypotheses:

**H1: There is a significant negative relationship between the stringency of lockdown measures and the number of startups in China during the COVID-19 pandemic.**

Hypothesis 1 (H1) posits that the imposition of stringent lockdown measures by the government to curb the spread of COVID-19 has had an adverse effect on the entrepreneurial ecosystem. The restrictions on movement, social interactions, and business operations may have deterred potential entrepreneurs from initiating new businesses, resulting in a decline in the number of startups.

*2.2 Hypothesis Development of the Impact of COVID-19 on Local Government Public Finance*

The hypothesis H2 of this study is that the Covid-19 pandemic has had a negative impact on the public finance of local governments in China. This hypothesis is based on the assumption that the containment measures taken to control the spread of the virus have directly or indirectly affected the revenue-generating capacity of local governments, leading to a decline in fiscal revenue. Additionally, the increased need for public health spending and economic support measures during the pandemic is expected to have resulted in an increase in local government expenditure.

**H2: The Covid-19 pandemic has negatively affected the public finance of local governments in China.**
**H3: The Covid-19 pandemic has positively affected the public finance of local governments in China.**

To test H2 and H3, the study will employ a two-way fixed effects regression model, using data from 39 Chinese cities on a monthly basis. The model will control for potential confounding factors such as CPI and local house prices.

*3. Data and Research Design*

*3.1.1 Sample Selection of the Impact of Lockdown on Startups*

This study employs monthly province-level data from 2020 to 2022, encompassing all 31 provinces in China, and comprises a total of 1,116 observations.

The number of newly registered businesses is adopted as the primary measure of startups in this study, whereby newly registered businesses refer to those registered with the administrative department for industry and commerce in accordance with national laws, regulations, and relevant provisions. The monthly number of newly registered businesses collected illustrates a steady increase over time. All data utilized in this study are monthly provincial data, sourced from the Wind dataset.

To determine the stringency of lockdown measures in each province, this study uses data from the Oxford COVID-19 Government Response Tracker (OxCGRT) program, which comprises 23 indicators encompassing school closures, travel restrictions, and vaccination policy. The average stringency index is chosen as the indicator of lockdown measures, which ranges from 0 to 100, with higher values denoting stricter lockdown measures. Notably, the lockdown policy was first implemented in Wuhan in January 2020 and was gradually extended to all 31 provinces included in this study, with the country's government declaring the lifting of COVID-19 lockdown in December 2022.

*3.1.2 Sample Selection of the Impact of COVID-19 on Local Government Public Finance*

This section focuses on studying the impact of the Covid-19 epidemic on fiscal revenues, fiscal expenditures, and government debt of thirty cities, using monthly data. The epidemic intensity indicator is measured by the monthly number of confirmed patients in the relevant cities, while macro indicators such as CPI and local house prices are considered as control variables. The data sources include the Finance Bureau of each city, the National Health Commission of the People's Republic of China, and the National Bureau of Statistics of China.

It should be noted that fiscal data for January of each year are excluded from this study because fiscal data are cumulative, and some cities publish data to unify January and February as February releases (important point).

In terms of indicator selection, this section uses the number of diagnosed patients in cities to measure the local epidemic risk level. Additionally, due to differences in price levels in different places and the fact that land concessions account for a large proportion of local government finances, this section also uses the house prices of each city as an explanatory variable.

In this section, all the data are monthly. The data on local government public finance, CPI, and GDP are sourced from the Wind dataset. As fiscal data for local governments are typically collected monthly, some local governments may have missing data for the month of January. Therefore, to ensure data consistency, this paper excludes all January data from the analysis. The data on the number of confirmed patients in each city is sourced from the Covid-19 record program.

Table 1 Descriptive Statistics

| VARIABLES | (1) N | (2) mean | (3) sd | (4) min | (5) max |
|---|---|---|---|---|---|
| Public Expenditure | 1,703 | 973.4 | 1,272 | 416 | 8,431 |
| Public Revenue | 1,703 | 792.0 | 1,151 | 328 | 7,772 |
| CPI | 1,703 | 102.0 | 1.111 | 97.80 | 105.9 |
| House Price Index | 1,703 | 100.5 | 0.655 | 98.70 | 106.2 |
| ConfirmedCount_city | 1,703 | 389.4 | 3,966 | 0 | 50,431 |
| Number of ct | 29 | 29 | 29 | 29 | 29 |

For the variable public expenditure, the sample mean is 973.45 with a standard deviation of 1272.13, a minimum value of 416, and a maximum value of 8430.9. For the variable public revenue, the sample mean is 792.01 with a standard deviation of 1150.81, a minimum value of 328.47, and a maximum value of 7771.8. For the variable CPI, the sample mean is 101.97 with a standard deviation of 1.11, a minimum value of 97.8, and a maximum value of 105.9. For the number of confirmed patients, the sample mean is389.4 with a standard deviation of3966, a minimum value of 0, and a maximum value of 50431.

*3.2.1 Research Design of the Impact of Lockdown on Startups*

To study the effect of COVID-19 on startups, we use difference-in-difference as the identification strategy and estimate the following regression:

$$\text{Startups}_{it} = \alpha + \beta_1 \text{Stringency}_{it} * \text{Province}_i + \beta_2 \text{Province}_i + \beta_3 \text{Time}_t + \varepsilon_{it},$$

where Startups$_{it}$ is the number of newly registered enterprises in province i in time t. The Stringency$_{it}$ is the average value of Oxford stringency index of province i in time t. Province$_i$ is the fixed effect of province which can control the different characteristics of each province such as entrepreneurial environment and innovative culture. Time$_t$ is the fixed effect of time.

We assumes the presence of multiple fixed effects, including province, month, and year fixed effects. Controlling for these fixed effects allows for the estimation of the effect of lockdown measures on the number of startups. A significant estimate of the coefficient of interest, as determined through a t-statistic test, indicates a significant causal relationship between COVID-19 lockdown measures and startups. Conversely, a non-significant estimate suggests no causal relationship between lockdown measures and startups.

*3.2.2 Research Design of the Impact of COVID-19 on Local Government Public Finance*

In this paper, we investigate the impact of the epidemic on local government finances using a two-way fixed effects regression on the interactive term of the number of cities and confirmed patients. The model is as follows. We should also consider the pre-trend (parallel trend assumption):

$$Y_{it} = \alpha + \gamma_i + \delta_t + \beta N_{it} + \theta X + \varepsilon_{it}$$

This model includes several variables to examine the impact of local government policies on the number of confirmed COVID-19 cases. The intercept is represented by α, while $\gamma_i$ represents the unit fixed effect and $\delta_t$ represents the time fixed effect. Of particular interest is the local government epidemic impact coefficient, β. This coefficient measures the direct impact of local government policies on the number of confirmed diagnoses. Other control variables, including CPI and house prices, are represented by X, with their respective coefficients of control variables represented by θ. Ultimately, the model aims to assess the impact of local government policies while controlling for other potentially confounding variables.

*4. Results*

*4.1 Results of the Impact of Lockdown on Startups*

This section proposes a model to empirically study the relationship between the scale of newly registered businesses and lockdown stringency, CPI, and GDP. The results demonstrate a strong correlation between newly registered businesses and lockdown stringency. Table 2 presents the descriptive statistics of each variable. The large standard deviation indicates that newly registered businesses and commercial economic activities experienced significant fluctuations during the pandemic period, impacted by regions and seasons. Table 3 displays the regression results, which fit well with the actual data. Notably, the stringency variable demonstrates statistical significance. Furthermore, to investigate whether the stringency index of lockdown has a hysteresis effect on the number of newly registered enterprises, we will employ a regression analysis using a first-order lag variable $L\_Stringency_{i,t-1}$:

$$Startups_{it} = \alpha + \beta_1 Stringency_{it} * Province_i + \beta_2 L\_Stringency_{it-1} + \beta_3 Province_i + \beta_4 Time_t + \varepsilon_{it},$$

the result of table 4 shows that the first order lag variable does not significantly affect the number of startups, indicating that there is no hysteresis effect of stringency index.

Table 2

Mean, Standard Deviation, Minimum and Maximum of Startup and Stringency Index

| Variable | Obs | Mean | Std.dev. | Min | Max |
| --- | --- | --- | --- | --- | --- |
| startup | 1,116 | 23726.30 | 22090.95 | 89 | 138713 |
| stringency | 1,116 | 54.91 | 13.56 | 14.20897 | 93.85036 |

Table 3

Effect of Stringency Index on Startup without Lag Variable

| startup | coefficient | Std. Err. | t | P>|t| | [95% cond. Interval] | |
|---|---|---|---|---|---|---|
| stringency | -116 632 | 43 45002 | -2.68 | 0.012 | -205 3687 | -27.89519 |
| _cons | 30130.78 | 2385.921 | 12.63 | 0.000 | 25258.08 | 35003 48 |

Table 4

Effect of Stringency Index and First-Order Lag Variable of Stringency Index on Startup

| startup | coefficient | Std. Err. | t | P>|t| | [95% cond. Interval] | |
|---|---|---|---|---|---|---|
| stringency | -101.921 | 41.38041 | -2.46 | 0.020 | -186.4311 | -17.41098 |
| l_stringency | 25.70289 | 30.73945 | -0.84 | 0.410 | -88.48122 | 37.07544 |
| _cons | 31127.78 | 2835.084 | 10.989 | 0.000 | 25337.77 | 36917.79 |

This study classified China's 31 provinces into seven geographical divisions according to China's geographical divisions. These divisions are Northeast of China, North of China, Northwest of China, Central China, South of China, Southwest of China, and East of China. Time series images of the number of new businesses in each region were generated to reflect the changes in newly registered businesses with the development of the pandemic. On average, all provinces had the lowest number of newly registered business applications in February of each year. Provinces with more developed economies, such as Beijing, Shanghai, Guangdong, Zhejiang, and Jiangsu, experienced a significant decrease in the number of newly registered businesses at the beginning of the COVID-19 outbreak, with an average reduction of 60%. In Hubei Province, the source of COVID-19 outbreaks in China, the number of newly registered businesses decreased by 98% in the early stages of the pandemic due to seasonal factors and the pandemic shock. These results suggest that the government's stringency lockdown policy and the sudden pandemic significantly impacted local commercial and outdoor activities, leading to negative effects on the economy.

After the initial wave of the pandemic, the number of newly registered businesses exhibited an overall increasing trend. This can be attributed to the implementation of mature lockdown policies and a targeted medical protection system in China, which enabled timely prevention and treatment of new COVID-19 cases. Furthermore, as various regions gradually resumed normal activities, business activities were able to pick up once again. Notably, the COVID-19 virus in 2020, the first year of the pandemic, did not undergo multiple rounds of mutations and its overall characteristics were based on severe symptoms. Furthermore, the transmission and incubation period of the virus were distinct from those of the Omicron variant and other more transmissible variants. This explains why there were no regional outbreaks of COVID-19 in China following the initial wave of the pandemic, and business activities were able to flourish under the effective control of the pandemic.

The Chinese government has consistently pursued a Zero-COVID policy, which has included implementing lockdown policies for specific areas based on the severity of the pandemic. This set of policies proved effective in the early and mid-term of epidemic prevention, improving pandemic prevention and control efficiency. However, with the mutating virus, regional lockdowns and Zero-COVID policies have become less effective. Once the number of cases and suspected cases exceeds a certain threshold, patients' primary activity areas are closed and controlled for a set period of time. Consequently, the number of newly registered businesses in various areas fluctuated significantly during the middle and late stages of the COVID-19 pandemic.

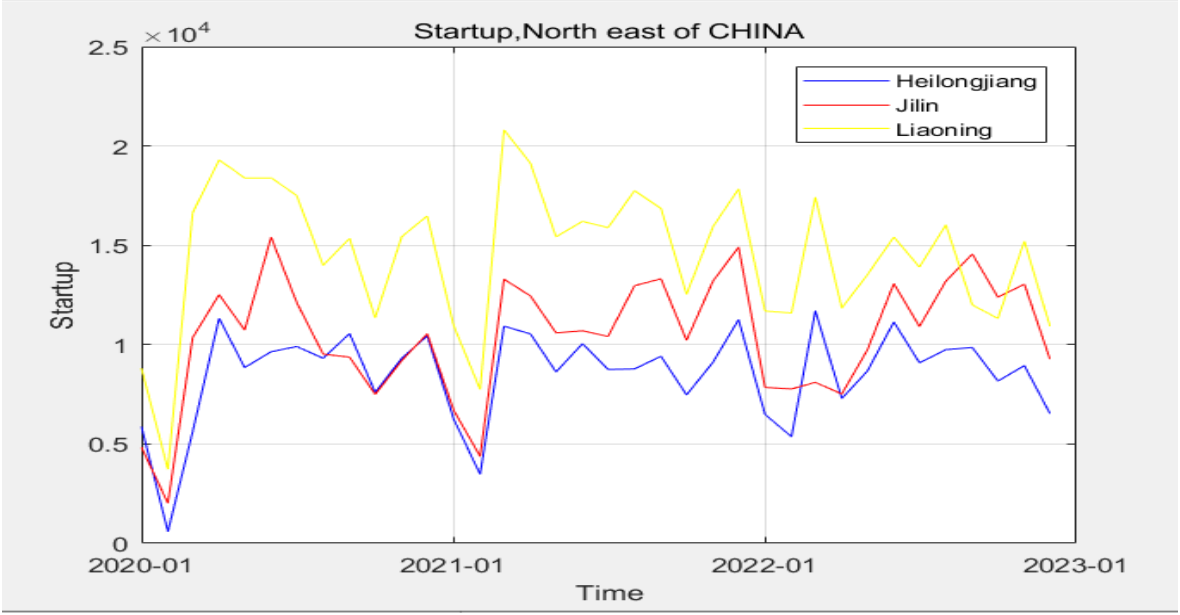

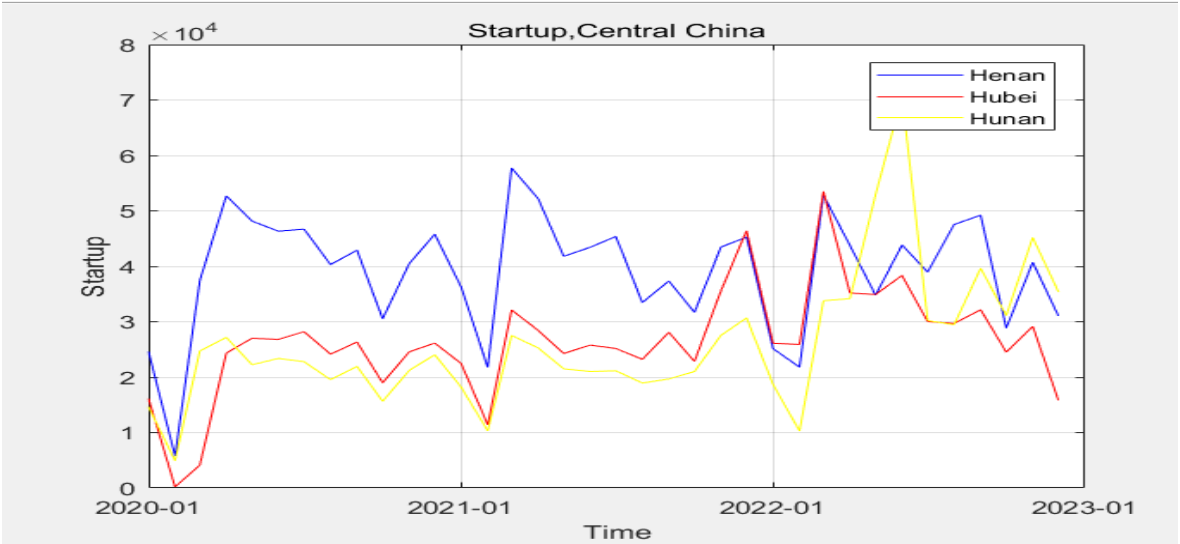

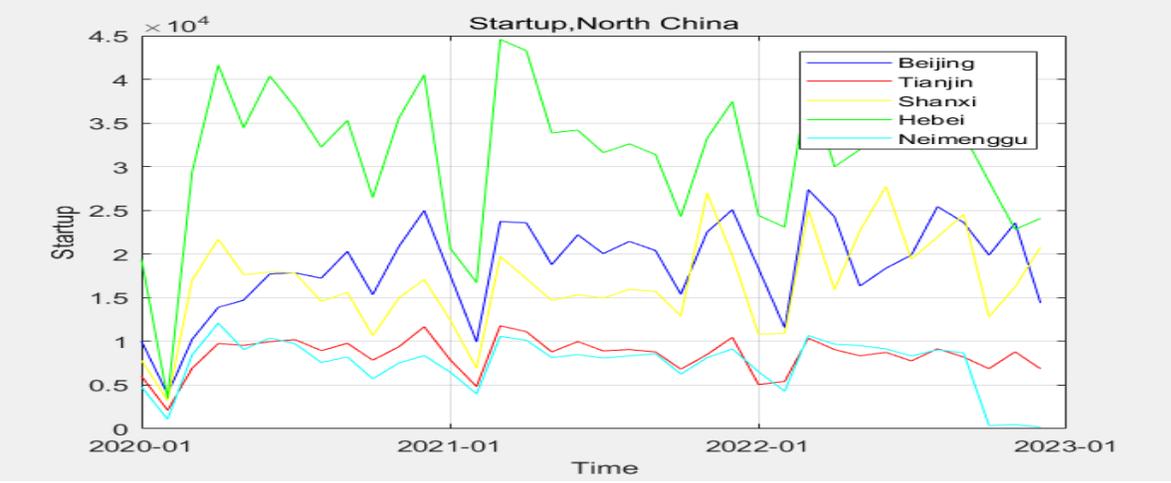
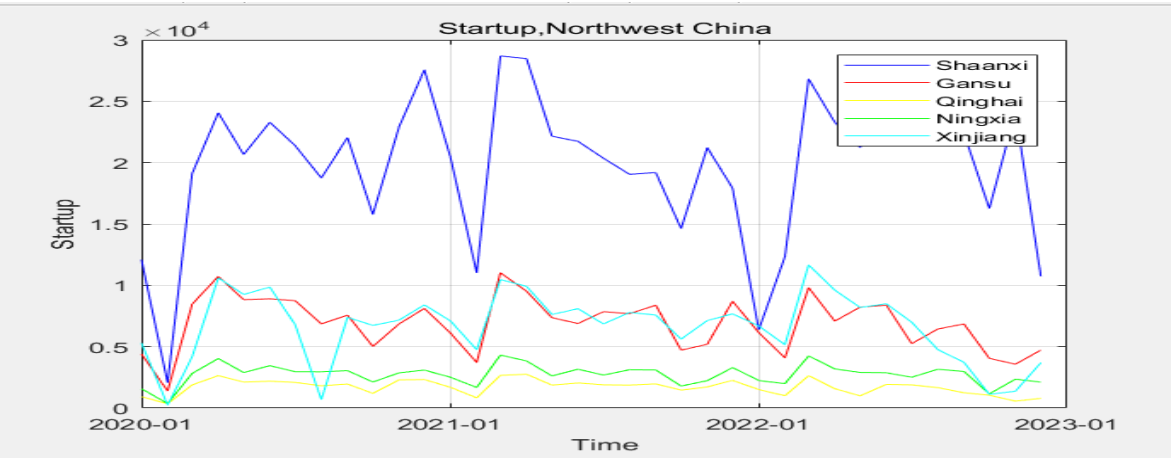

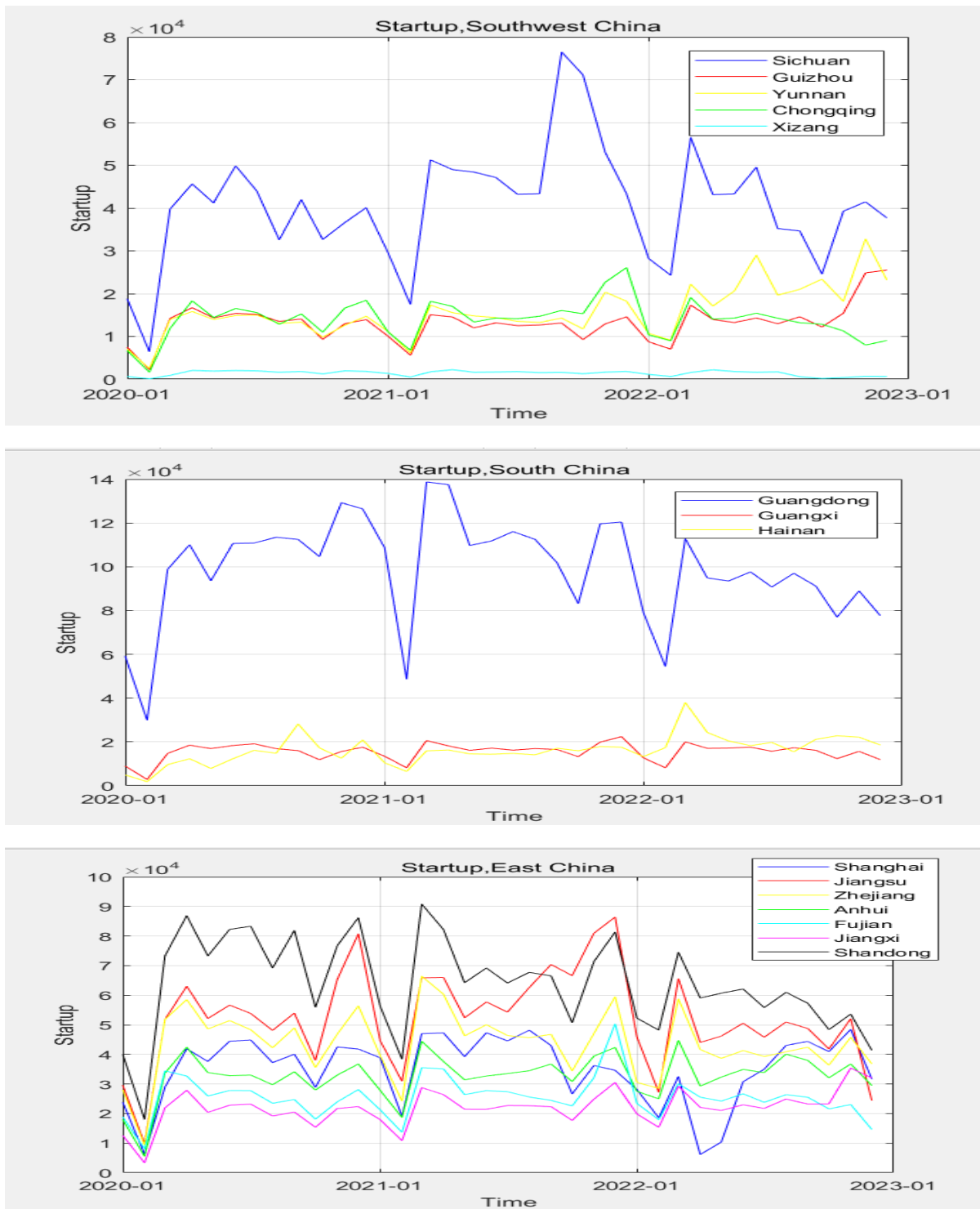

Figure 1 The Change of Number of Newly Registered Enterprises in each Region

*4.2 Results of the Impact of COVID-19 on Local Government Public Finance*

This model uses a two-way fixed effect to study the effect of COVID-19 on local government public finance, and the results demonstrate that the development of COVID-19 affects local government revenues. Table 4 illustrates that the regression model has a high degree of fit with the actual data, and the variable "number of confirmed patients" is statistically significant.

Table 5 Two-Way Fixed Effects Analysis

| VARIABLES | (1) Public Revenue | (2) Public Revenue | (3) Public Expenditure | (4) Public Expenditure |
|---|---|---|---|---|
| ConfirmedCount_city | -0.000 | -0.003*** | 0.002 | 0.002** |
|  | (-0.08) | (-1.73) | (0.56) | (1.84) |
| Public Expenditure |  | 0.789*** |  |  |
|  |  | (113.74) |  |  |
| CPI |  | 1.285 |  | -4.604 |
|  |  | (0.23) |  | (-0.70) |
| House Price Index |  | 17.053** |  | -19.486** |
|  |  | (2.54) |  | (-2.43) |
| Public Revenue |  |  |  | 1.127*** |
|  |  |  |  | (113.74) |
| Constant | 250.486*** | -1,767.363** | 225.026** | 2,361.721** |
|  | (3.20) | (-2.01) | (2.41) | (2.25) |
| Observations | 1,703 | 1,703 | 1,703 | 1,703 |
| R-squared | 0.389 | 0.933 | 0.466 | 0.941 |
| Number of cities | 29 | 29 | 29 | 29 |
| city FE | YES | YES | YES | YES |
| time FE | YES | YES | YES | YES |

t-statistics in parentheses
*** $p<0.01$, ** $p<0.05$, * $p<0.1$

According to the regression output, we can see that number of confirmed contracted covid-19 has a statistically significant negative coefficient of -0.003, indicating that an increase in confirmed COVID-19 cases in a city is associated with a decrease in local government fiscal revenue. This result demonstrates that the COVID-19 pandemic has a negative impact on local government fiscal revenue, potentially due to decreased economic activity or government spending to respond to the pandemic.

This regression result suggests that local government public expenditure is positively associated with local government public revenue and the number of patients confirmed contracted with Covid-19, while negatively associated with customer price index and house price index. As the variable we are interested in, the coefficient of number of confirmed covid-19 is 0.002127, which means that a one-unit increase in number of confirmed covid-19 is associated with a 0.002127 unit increase in public expenditure, all else being equal. This coefficient is statistically significant at the 0.05 level, as indicated by the low p-value (P>z=0.001). However, the coefficients for CPI and H_P are not statistically significant, indicating that their effects may be weaker or more uncertain.

*5. Conclusions*

The COVID-19 outbreak has had a significant impact on the public finance of Chinese local governments and the startup ecosystem in China. Containment measures taken by local governments, such as lockdowns and travel restrictions, have directly or indirectly affected their revenue capacity and led to an increase in fiscal pressure by 14.53% during the outbreak. The fiscal revenue and expenditure gap accounted for almost half of local fiscal revenue. Moreover, the economic downturn caused by the pandemic has led to a decline in tax revenue for local governments, further exacerbating their fiscal pressure. As a result, some local governments have turned to borrowing to finance their operations and investments, leading to an increase in local government debt.

In dealing with the public health crisis, local governments in China have had to increase their expenditures on investments in medical equipment, personnel, and infrastructure, as well as implement social welfare measures to support individuals and businesses affected by the outbreak. Concurrently, lockdown measures have negatively impacted entrepreneurship, with the average stringency index showing a significant negative effect on the number of startups. This decrease in the growth rate of newly registered companies may lead to risks of liquidity crunch, unemployment, and even economic recession.

Despite the existence of numerous incubators guided by the government, the quality of services provided by these incubators needs enhancement, and the enthusiasm for entrepreneurship has been dampened during the pandemic. The government should be more concerned about the survival and development of startups, implementing more preferential policies focused on them. Long-term measures should be embedded in and supported by the wider entrepreneurial ecosystem to ensure rapid recovery and growth (Kuckertz et al, 2020). Furthermore, financial institutions should fulfill their social responsibilities by lending not only to large enterprises but also paying attention to the growth of startups.

Overall, the COVID-19 outbreak has had a significant impact on the public finance of Chinese local governments and the startup ecosystem, with both revenue and expenditure being affected, as well as the growth and development of new businesses.